# Pivotuner: automatic real-time pure intonation and microtonal modulation

Dmitri Volkov
Indiana University Bloomington
dvolkov@iu.edu

## ABSTRACT

*Pivotuner is a VST3/AU MIDI effect plugin that automatically tunes note data in an adaptive pure intonation, in real time. Where previously pure intonation was out of reach for most musicians due to difficulty and impracticality, Pivotuner enables it to be achieved easily and straightforwardly by using novel yet simple algorithms. This may lead to more widespread exploration of pure intonation for a larger and more diverse crowd of musicians!*

*This paper includes a review of prior systems for adaptive pure intonation systems, including Hermode Tuning[1]/Kontakt Dynamic Pure Tuning[2] and Just Intonation[3]. The paper introduces the notion of an adaptive tuning center and how it serves as a flexible underlying concept for multiple tuning algorithms, as well as extensions to offer greater control for performers, including pitch and tuning center locking and resetting, and gradual interpolation between equal temperament and pure intonation. The paper then showcases some pieces which use Pivotuner effectively, then discusses areas for future exploration within Pivotuner's feature set, and plans for future development.*

## INTRODUCTION

Pure intonation (also commonly referred to as "just intonation"), is loosely defined as tuning notes such that the ratio of their frequencies is simpler. Pure intonation is often considered to be ideal due to sounding more "in-tune" to the ear, as well as for having greater resonance and less beating[4]. However, pure intonation is seldom used in practice, partially because of the difficulty of aurally reproducing these frequency ratios and the inflexibility of fixed-pitch instruments (such as fretted and keyed instruments). While fixed-pitch instruments can be tuned in pure intonation, the limited set of notes means that while some chords sound highly resonant, most do not. This high presence of dissonances in pure intonation is what lead to the historical development of temperaments. While temperaments produce tunings that are subjectively enjoyable for a wide selection of chords, they are not pure intonation, and lack pure intonation's resonance and clarity.

This paper introduces Pivotuner, a VST3/AU audio plugin that tunes MIDI[14] data in pure intonation, using a customizable set of algorithms. Pivotuner enables pure intonation to be achieved on (electronic) keyboard instruments. While prior work exists on using computers to realize pure intonation, Pivotuner enables a more advanced level of control and flexibility. The paper introduces the underlying principles of Pivotuner and outline how they work together to form a coherent and powerful system for real-time adaptive pure intonation. The paper presents two examples that make use of Pivotuner which otherwise would have been highly impractical to create, and by introducing areas for further exploration and development with Pivotuner.

## PRIOR WORK

This section explores prior work on the subject of achieving pure intonation with computers, and limitations of these systems.

### 1.1 Hermode Tuning/Kontakt Dynamic Pure Tuning

Hermode Tuning (HMT), is one of the leading automatic pure intonation systems, in partially because of its integrations with notable pieces of hardware and software, including the Logic Pro and Cubase digital audio workstations[5]. A similar system is included with the Kontakt sampler, as the Dynamic Pure Tuning system. While the two systems are different, they are similar enough such that only HMT will be discussed here.

HMT works by analyzing the active notes to discern intervals of fifths or thirds (or optionally sevenths), then uses pure or near-pure interval sizes to tune these notes[1]. Because these tunings may raise some notes and lower others relative to equal temperament, HMT attempts to raise and lower every note such that the net sum of raisings and lowerings is 0, in an attempt to stay close to equal temperament.

An advantage of HMT is that it produces convincing results while remaining near equal temperament, which makes it easy and intuitive to use for most musicians, who are primarily familiar with equal temperament. However, HMT is not very customizable: if a different tuning than the one HMT produces is desired, there is little flexibility. Furthermore, HMT is explicitly designed to avoid microtonal modulation, which limits musicians who may desire to use it in a piece.

### 1.2 Just Intonation

The confusingly named Just Intonation is an app that models pure intonation as a linear system of equations that, when solved, provides information about the optimal tuning[3]. By attempting to find a tuning that produces near-pure ratios between as many notes as possible, Just Intonation is a very flexible system, and can produce convincing results for many chords. When tuning, it also takes previous note data into account to ensure a more consistent tuning over time. While by default Just Intonation avoids microtonal modulation, it can be enabled. Furthermore, Just Intonation supports customizable intervals.

While the linear algebra-based approach of Just Intonation is very powerful, it does suffer from some limitations. One is that for dense chords, it tends to decay back into equal temperament-based tuning. While this is mitigated somewhat by the ability to weight intervals by importance, so that e.g. consonant intervals are tuned purely at the expense of the pure tuning of dissonant intervals, it offers little consolation for musicians who desire pure tunings of dense chords. Furthermore, while there is flexibility that microtonal modulation can be enabled or disabled, it is difficult to enable/disable it in real time for cases where a musician may want to use microtonal modulation selectively. Finally, Just Intonation is only available as a standalone app, which makes it difficult to use in conjunction with other software.

### 1.3 Other Approaches

There are other attempts to realize adaptive pure intonation, such as Mutabor[6] and the Groven Piano[7]. However, because of technical and practical constraints, it was not possible to review these at the time of writing; they are included here for completeness.

## OVERVIEW OF PIVOTUNER

Pivotuner is a VST3/AU MIDI effect plugin that translates single-channel MIDI note data into purely-tuned MPE[15] data. The general process starts with determining a "key" note, using one of several user-configurable algorithms. Pivotuner then tunes the rest of the active notes relative to the key note by using one of several customizable tuning algorithms. Besides flexibility regarding which algorithm is used for key and tuning selection, Pivotuner also offers methods to customize the tuning of each interval and chord, to control when a new key is selected, to and to control microtonal modulation.

### 2.1 Key Determiner Algorithms

When Pivotuner receives MIDI data, it uses a Key Determiner algorithm to determine a note relative to which all other notes should be tuned. As of writing, Pivotuner supports three Key Determiners: Length Key Determiner, Bass Key Determiner, and Chord Key Determiner.

Length Key Determiner sets the key to be whichever note has been held down the longest, which is the same as whichever note was played first and maintains to be held down. For example arpeggiating a root position C triad chord upward (while keeping the notes held down) would tune to C, while doing the same downward would tune to G, because those are the notes which would have been played the longest. This method of key selection is very powerful because a performer can choose any note as a key simply by playing it! If a performer wants to maintain a key without keeping the note held down, they can use the Key Lock, which is described later.

Bass Key Determiner sets the key to be the lowest note which is active. It generally behaves similarly to the Length Key Determiner, but offers less control. However, it may be more intuitive for some musicians, and there are other cases where it may be preferable.

Chord Key Determiner chooses the key by attempting to match the active notes with a chord in a user-customizable database. If such a match is made, the key is set to the root of the detected chord, otherwise the Chord Key Determiner fails to determine a key, causing Pivotuner to default to equal temperament. Chord Key Determiner is designed to work with Chord Tuning, which is described later.

### 2.2 Tuning Algorithms

Once a key is determined, Pivotuner uses a Tuning algorithm to figure out how to tune the active notes. As of writing, Pivotuner supports four tuning algorithms: Scale Tuning, Mirror Scale Tuning, Interval Tuning, and Chord Tuning.

Scale Tuning is based on traditional pure intonation scales. Scale Tuning uses the user-specified tunings for intervals from minor 2nd to major 7th as tunings for each note in a chromatic scale with a key center of the active Pivotuner "key." Without Pivotuner's adaptive key centers, this would cause Pivotuner to produce the same results as an untempered keyboard.

Mirror Scale Tuning is similar to Scale Tuning, except that notes below the key note are tuned according to their inverted interval away from the key note. For example, assuming a 7:4 tuning for the minor 7th and a 9:8 tuning for the major 2nd, when C4 is the key, then Bb5 would be tuned in a 7:4 ratio to C, whereas Bb4 would be tuned in a 8:9 ratio. Mirror Scale Tuning makes it much easier to use the harmonic tuning of a minor 7th (7:4), as going down a step from the key sounds like a normal 9:8 step, instead of the wide septimal 8:7 step.

Interval Scale Tuning tunes the intervals of adjacent notes purely, starting with the root. For example, a chord consisting of C4, G4, and D5 played with C4 as the key and 3:2 as the perfect 5th interval, would tune the G in a 3:2 interval to the C, and the D in a 3:2 interval to the already tuned G. If G was the key, then C would be tuned in a 2:3 interval to the G, and the D would be tuned in a 3:2 interval to the G. This algorithm can tend to produce dissonant tunings as octaves of the same note tend not to be in tune with each other.

Chord Tuning uses the same database of chords as the Chord Key Determiner. This database, which can be edited by the user, contains a list of chords with a specified root and tunings for each note in the chord. Chord Tuning attempts to match the played note to a chord and if it finds one, tune the notes according to the specified tuning. If no chord is found, it defaults to equal temperament.

### 2.3 Microtonal Modulation

Microtonal modulation in Pivotuner occurs naturally with the established systems in place: as long as non-equal-temperament intervals are specified, a microtonal modulation occurs whenever the Key Determiner reports a new key (and the Pitch and Key Locks, which are described later, are not enabled). Consider the following: a musician using Pivotuner with the Length Key Determiner and Scale Tuning plays a C, establishing the current key as C. With the C held down, they play an Eb, which is tuned 15.64 cents higher than it would be in equal temperament in order to create a pure 6:5 minor 3rd interval with the C. The C is then released, causing the Eb to become the key. Because the Eb, which was tuned 15.64c higher, is now the key, all notes will now be tuned relative to this 15.64c high Eb. Therefore a microtonal modulation up 15.64c has occurred! Microtonal modulations such as this can be chained to modulate farther; if the user were to change the key up a minor 3rd again, the net modulation would be up 31.28c. Furthermore, this type of microtonal modulation is fairly easy and intuitive to perform: just change keys to a note which is tuned in the desired modulation direction!

### 2.4 Key and Pitch Locking

Sometimes a musician may find themselves wanting to stay with a single consistent tuning, or otherwise avoid microtonal modulation. Pivotuner enables this through it's Key Lock and Pitch Lock features.

Key Lock prevents the key from changing when the Key Determiner reports a new key. Enabling the Key Lock can be useful for times when a musicians wants to stay within a single tuning system, or when they want to use a chord which does not have the key note in it.

Pitch Lock prevents microtonal modulation from occurring when the key changes. Normally when the key changes, the pitch deviation of the new key note from equal temperament remains the same as what is was before the microtonal modulation. With Pitch Lock enabled, the pitch deviation of the new key note will be the same as the pitch deviation of what the prior key note was, which effectively prevents any larger pitch change from occurring.

Both Pitch and Key lock can be mapped to MIDI Continuous Control parameters (CCs) such as the sustain pedal or modulation wheel, making them easy to toggle in live settings!

### 2.5 Key and Pitch Resetting

The current key and pitch can both be reset to user-specified default values. Resetting can be used to undo microtonal modulation, or to force equal temperament tuning. These are also controllable by MIDI CCs for convenience.

### 2.6 Bendback

The Bendback feature offers interpolation between the current purely-tuned chords and a completely unturned equal temperament. Bendback can serve both to enable back-to-back comparisons of tuned and untuned chords, and to enable more gradual resets. Bendback is controlled by a configurable MIDI CC.

## MUSICAL EXAMPLES

Below showcases some musical uses of Pivotuner by various artists, and the benefits of Pivotuner in each case.

### 3.1 IJ Pivotuner Jamming

Israel Strom, a popular Instagram jazz keyboardist known for his *Polytonal Polyrhythm Warmup* video[8], and soon to go on tour with the drummer Louis Cole, recorded an improvisation using Pivotuner, called *Pivotuner Jamming*[13], that incorporates microtonal modulations into a more familiar tonal language. This improvisation tends to stick to a key using key locking, then modulates around using pentatonic scales to chain whole step- and minor 3rd-based microtonal modulations. This jam demonstrates how Pivotuner can effectively be used with faster music to create a feeling of movement and energy. Furthermore, the microtonal modulations give a subtle yet powerful feeling of subtle fluctuation, that could be difficult to otherwise achieve, which heightens the experience and enjoyability of this jam.

The recording of *IJ Pivotuner Jamming* can be accessed at the following link:

`https://www.dmitrivolkov.com/projects/pivotuner/ij_pivotuner_jamming.mp3`

## 3.2 Xenexhibition

*Xenexhibition*[9] is a piece written explicitly to showcase Pivotuner in a musical context. *Xenexhibition* features three movements which are intended to respectively showcase: Pivotuner's ability to enable microtonal modulation and pure tuning of dense chords, technical consequences and effects of using Pivotuner, and how these can be used to heighten the emotional impact of a piece of music.

The first movement uses a theme based on one by Giambattista Benedetti, a mathematician from the 1500s who was the first to discover microtonal modulation[10]. It first demonstrates microtonal modulation by slightly shifting the pitch of a single note, before continuing to use the theme as a basic pattern for increasingly dissonant chords. It also demonstrates how a chord which is consonant in one key's pure tuning may be dissonant in another's.

The second movement is highly dissonant and aleatoric, and uses Pivotuner's microtonal modulation ability to the extreme. It uses large repeated scales and glissandos to quickly microtonally modulate vast distances, mostly through half- and whole-step-based microtonal modulations. The size of these modulations causes the sampler to dramatically retune notes, which heavily effects their timbre in fascinating ways.

The third and final movement reintroduces the theme from the first movement with the changed timbre from the second movement. A prominent feature of this movement is repeating the chord with a changing key, which subtly shifts the tuning and timbre of the chord.

A recording of Xenexhibition can be accessed at the following link:
`https://www.dmitrivolkov.com/projects/Xenexhibition.mp3`

## FUTURE WORK

Pivotuner is still a very young piece of software, with the first public release only in December 2022. As such, there is very much room for further exploration.

The possibilities of Pivotuner's existing feature set have not yet been fully explored. Pivotuner supports using any tuning for intervals from minor 2nd through major 7th, yet so far only ratio-based pure tunings for these intervals have been used. Therefore one area of further exploration is using alternate tunings for these intervals.

Pivotuner also not yet been used alongside acoustic instruments or voice in live settings. While it is difficult to perform microtonal music on acoustic instruments or with voice, Pivotuner may be able to provide microtonal reference pitches in a musical context, which may enable such performances to take place. Pivotuner may also be better equipped to accompany harmonic-series based instruments and vocal techniques such as overtone singing.

There are also plans to extend Pivotuner's feature set. One fairly basic extension would be to support MTS-ESP[14] and MIDI 2[16], to enable Pivotuner to work with a wider body of existing software. Chord Tuning can also be further improved such that different inversions of the same chord may be tuned differently, which is not currently the case. Furthermore, there are times when a musician may want to use different tunings for the same interval, e.g. septimal vs pythagorean tunings for the minor 7 interval. While Pivotuner does support customizable intervals, the intervals cannot be instantaneously switched, which makes it impractical for live performance settings. There also may exist undiscovered Key Determiner and Tuning algorithms that offer greater flexibility and power to musicians than the ones currently implemented in Pivotuner.

## CONCLUSIONS

Pivotuner enables a new level of power and control for musicians who wish to incorporate pure intonation and microtonal modulation. With a high degree of flexibility, Pivotuner can make it possible to consistency and reliably perform microtonal modulations in settings where it previously might not have been possible, such as live performance. Pivotuner has been used effectively in several works, but there is still very much room for further exploration, both in the use of Pivotuner and within the Pivotuner plugin itself.

For more information, please visit the Pivotuner website at:
`https://www.dmitrivolkov.com/projects/pivotuner/`
or reach out to the author!

**Acknowledgments**

The author would like to thank Ben Bloomberg for his help and mentorship in creating Pivotuner, and Jacob Collier for inspiring the creation of Pivotuner with his works *In the Bleak Midwinter*[11] and *Moon River*[12], as well as for his feedback on Pivotuner. For their help on *Xenexhibition*, thanks are due to Don Freund for his composition mentorship, John Gibson for providing hardware and support, and Erin Blake for the incredible performance! Furthermore, the author would like to thank Rayda Krell for her help with writing style and editing, and Minho Kang for reading an early version of this paper. The author would also like to thank the many other musicians who have tried out and provided feedback on Pivotuner; without them, Pivotuner would have never have developed as far as it has.